\documentclass{article}
\usepackage{spconf,amsmath,graphicx,tikz}
\usepackage{algorithm}
\usepackage{hyperref}
\usepackage{xcolor}
\usepackage{float}
\usepackage{optidef}
\usepackage{cases}
\usepackage[mode=build]{standalone}
\usepackage{amsmath}
\usepackage{amsfonts}
\usepackage{stmaryrd}
\usepackage{amsthm}
\usepackage[letterpaper, left=19mm,top=25mm, right=0.60in, bottom=0.60in]{geometry}
\usepackage{mathtools, nccmath}
\usepackage{graphicx}
\usepackage{caption}
\usetikzlibrary{automata, arrows.meta, positioning}
\usepackage[nodisplayskipstretch]{setspace}
\expandafter\def\expandafter\normalsize\expandafter{%
    \normalsize%
    \setlength\abovedisplayskip{2.5pt}%
    \setlength\belowdisplayskip{2.5pt}%
    \setlength\abovedisplayshortskip{2.5pt}%
    \setlength\belowdisplayshortskip{2.5pt}%
}

\title{Implicit Neural Multiple Description for DNA-based data storage}

\name{Trung Hieu Le$^*$, Xavier Pic$^*$ \thanks{$^*$These authors contributed equally to this work}, Jeremy Mateos, Marc Antonini}
\address{I3S laboratory, Côte d’Azur University and CNRS, UMR 7271, Sophia Antipolis, France}

\begin{document}
\maketitle
\begin{abstract}
DNA exhibits remarkable potential as a data storage solution due to its impressive storage density and long-term stability, stemming from its inherent biomolecular structure. However, developing this novel medium comes with its own set of challenges, particularly in addressing errors arising from storage and biological manipulations. These challenges are further conditioned by the structural constraints of DNA sequences and cost considerations.
In response to these limitations, we have pioneered a novel compression scheme and a cutting-edge Multiple Description Coding (MDC) technique utilizing neural networks for DNA data storage. Our MDC method introduces an innovative approach to encoding data into DNA, specifically designed to withstand errors effectively.
Notably, our new compression scheme overperforms classic image compression methods for DNA-data storage. Furthermore, our approach exhibits superiority over conventional MDC methods reliant on auto-encoders. Its distinctive strengths lie in its ability to bypass the need for extensive model training and its enhanced adaptability for fine-tuning redundancy levels.
Experimental results demonstrate that our solution competes favorably with the latest DNA data storage methods in the field, offering superior compression rates and robust noise resilience.
\end{abstract}


\vspace{-0.75\baselineskip}
\section{Introduction}
\vspace{-0.75\baselineskip}
The memory of humanity hinges on our capacity to effectively handle ever-expanding volumes of data, spanning timeframes ranging from mere years to several centuries. As our current storage media struggle to keep pace, there is an urgent need to explore groundbreaking solutions that can be swiftly put into practical use. In the development of alternative data storage methods, synthetic molecules, particularly synthetic DNA, appear as one of the most promising options. Due to its density, durability, and its low energy consumption, synthetic DNA is an ideal storage support candidate for long-term data storage.
The initial phase in the data encoding process involves constructing a sequence of nucleotides A, T, C, and G (referred to as nts). However, it is imperative that the DNA-encoded information stream follows specific biochemical constraints. These constraints include avoiding homopolymers, maintaining a balanced GC content, and preventing repetitive patterns. Additionally, it is crucial to acknowledge that the biochemical procedures involved in this process can introduce errors that may compromise the integrity of the stored data. Operations such as synthesis, sequencing, storage, and DNA manipulation can introduce errors in the form of substitutions and indels (insertions or deletions of nucleotides).
During the last decade, information therorists have developed different schemes for the encoding of digital data into DNA, with some of them targeting the storage of images \cite{DNARAM,Dimopoulou2021AJI}. Some compression algorithm and coders have been developed specifically for this paradigm of data storage \cite{Church, Goldman2013, Aeon, MELPO}.
This work introduces a Single Description Coder (SDC) and a Multiple Description Coder (MDC) designed for DNA data storage with the SDC method exhibiting superior compression performance compared to the existing state of the art. 
\begin{figure*}[!htp]
    \centering
    \includegraphics[width=0.80\linewidth]{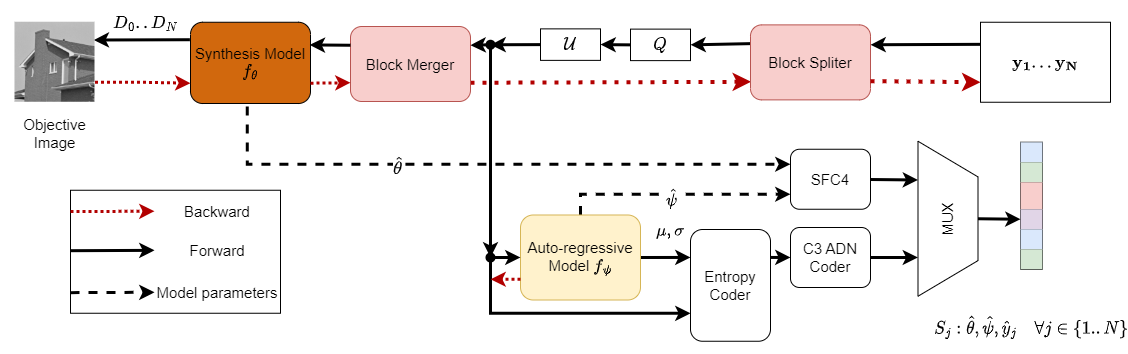}
    \vspace*{-0.5\baselineskip}
    \caption{\footnotesize
    \textbf{SF-MDC}: During the training process, $N$ latent sets are initially divided into blocks of size 8x8. Each block is then quantized independently with added uniform noise. These quantized latent blocks are then fed into the Block Merger module. In this module, each block is categorized as either redundancy or principal. Principal blocks are merged to form the central description, as illustrated in Fig. \ref{fig:robinrounditem}. Both these side descriptions and the central description are then input into the synthesis module, which generates the corresponding reconstruction and computes the related distortion. The latent space is updated using the back-propagation process, which is based on the distortion measured in MSE. Simultaneously, the Auto-regressive model is refined to better estimate the distribution of the quantized latent space.}
    \label{fig:globalarch}
    \vspace*{-\baselineskip}
\end{figure*}
MDC for image encoding involves encoding multiple representations of an image; if one is lost or corrupted during transmission, the remaining descriptions can still be used to reconstruct the original image with some quality degradation. Recent research \cite{DBLP:journals/tcsv/ZhaoBWZ19,DBLP:journals/mta/ZhaoZBWZ22} show a potential use of neural networks to generate different descriptions, which involve Generative Networks and Compressive Autoencoders. 
However, the main drawback of this method is its long training process that has a high computational cost.
Furthermore, the training process must be performed with very large datasets to converge towards an optimal model. This is even more challenging in the MDC context due to the redundancy adaptation mechanism, which requires retraining the model.

In recent works on image compression using neural networks, the so-called Implicit Neural Representation (INR), learns to represent an image implicitly through its weights, a coordinate map, and possibly a latent space \cite{sitzmann2019siren,DBLP:conf/eccv/StrumplerPYGT22}. More recently, the Coordinate-based Low Complexity Hierarchical Image Codec (COOL-CHIC) framework \cite{arxiv:ladune2023coolchic} has achieved superior performance compared to traditional image compression methods. The first MD scheme using INR (INR-MDSQC) is proposed in \cite{le2023inrmdsqc} with the following advantages: generalized model training is unnecessary, high performance and flexible redundancy tuning. However, INR-MDSQC's drawback is the number of descriptions, which is limited to two. Moreover, those descriptions are not balanced. 
The goals of implementing Multiple Description Coding (MDC) in DNA data storage are twofold: minimizing the reading cost, and enhancing noise robustness. This is particularly crucial due to the biochemical constraints inherent in the process, which can result in the absence of certain oligos.
To our knowledge, this work constitutes the first MDC application for DNA data storage. More precisely,  we propose a Spatial Frequency Multiple Description based on INR (SF-MDC) generalized to $N$ descriptions, and evaluate its performance on the Kodak Lossless True Color Image Suite dataset.\footnote{\color{blue}\url{http://r0k.us/graphics/kodak/}}.
\vspace*{-1.0\baselineskip}
\section{Spatial frequency MDC \label{sec:MDCpbstatement}}
\vspace*{-0.5\baselineskip}
In this section, we introduce a SF-MDC approach that incorporating an INR. The SF-MDC architecture, as depicted in Fig. \ref{fig:globalarch}, comprises four main components:
\vspace*{-0.5\baselineskip}
\begin{itemize}
    \itemsep -0.3em
    \item $N$ sets of hierarchical latent spaces
    \item $f_\theta$: Synthesis Model with \(\theta\) its parameters
    \item $f_\psi$: Auto-regressive Model with \(\psi\) its parameters
    \item Block Splitter/Merger
\end{itemize}

\vspace*{-1.5\baselineskip}
\subsection{Synthesis model}
\begin{figure}[!htp]
\vspace*{-1.5\baselineskip}
    \centering
    \includegraphics[width=0.9\linewidth]{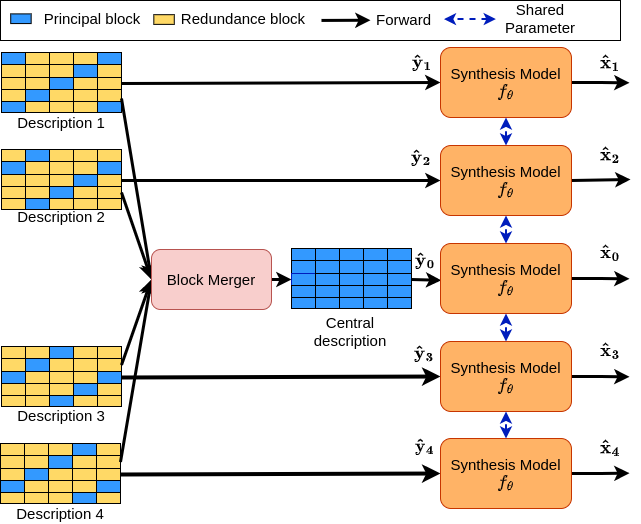}
    \caption{\footnotesize \textbf{Block Merger Module}: In this example, the number of descriptions is \(N=4\). The Principal and Redundancy blocks are assigned using the principle of round-robin item attribution. The central description is derived from the principal blocks of the 4 descriptions. Each description is then sequentially fed into the Synthesis model.}
    \vspace*{-1.5\baselineskip}
    \label{fig:robinrounditem}
\end{figure}

\vspace*{-0.5\baselineskip}
The quantization process is defined as follows:
\begin{equation}
\hat{s} = Q(s,\Delta s)
\label{eq:scalarquatizer}
\end{equation}
where $s$ is the element to be quantized, and $\Delta s$ is its associated quantization step. The latent spaces corresponding to each description $\mathbf{y}_j \in \{\mathbf{
y}_1..\mathbf{y}_N\}$ are hierarchically organized at different levels of resolution. Accordingly, we denote by $\mathbf{y}_{k|j}$ the latent space corresponding to resolution level $k$ for description $j$. 
In our solution, each description contains a mix of redundancy (low rate, low quality) and principal (high rate, high quality) blocks. At the decoder, when all the descriptions are received correctly, the decoder will merge all the principal blocks to form the central description. Otherwise, the redundancy blocks will be used to replace any corrupted principal blocks. Therefore at the encoding phase,
to distribute equal amounts of redundancy across descriptions, block splitter divides each $\mathbf{y}_{k|j}$ into $M$ blocks, each of size 8x8. The segment of the latent delineated by block $b$ is denoted as $y_{k|j}^{b}$, where $b \in \{0,1,...,M-1\}$ is the block index. Each $y_{k|j}^{b}$ is quantized with a unique quantization step $\Delta{y_{k|j}^{b}}$. A principal block uses a finer step, and a redundancy block uses a coarser step.
The central description $\mathbf{\hat{y}}_0$ is merged from the principal blocks as depicted in Fig. \ref{fig:robinrounditem}. Hence, each quantized block $\hat{y}_{k|j}^{b}$ is expressed as:
\begin{equation}
\hat{y}_{k|j}^{b} = Q(y_{k|j}^{b},\Delta y_{k|j}^{b})
\end{equation}
Therefore, the quantized latent space $\mathbf{\hat{y}_{k|j}}$ is defined as:
\begin{equation}
    \mathbf{\hat{y}}_{k|j} = \{\hat{y}_{k|j}^{b} \in \mathbb{Z}^{8\times 8}, b \in \{0,1,...,M-1\} \}
\end{equation}
From this, we deduce description $j$ composed by the set of different quantized latent spaces $\mathbf{\hat{y}}_{k|j}$ :
\begin{equation}
\mathbf{\hat{y}}_j = \{\mathbf{\hat{y}}_{k|j} \in \mathbb{Z}^{H_k\times W_k}, k \in \{0,1,..., L-1\}\}
\label{eq:hierarchic_def}
\end{equation}
where $H_k=\dfrac{H}{2^k}$, $W_k=\dfrac{W}{2^k}$, and $L$ denotes the hierarchical depth of $\mathbf{\hat{y}}_j$. As discussed in \cite{le2023inrmdsqc}, $\mathbf{\hat{y}}_{j}$ is sequentially input into the synthesis model $f_\theta$ with shared parameters, transforming set of latent spaces into a reconstructed image. In the synthesis model $f_\theta$, each $\mathbf{\hat{y}}_{k|j}$ is first upsampled using bi-cubic interpolation to match the target image shape before being fed into the MLP. The output of $f_\theta$ is defined as: 
\begin{align}
\mathbf{\hat{x}}_{j}&=f_\theta(\mathbf{\hat{y}}_j) \quad \text{where} \quad j\in\{0, .., N\}
\end{align}
The distortion of each $\mathbf{\hat{x}_{j}}$ compared to the target image is denoted by $D_j$ and measured using Mean Squared Error (MSE). Given that the latent space is discrete and the quantization process is non-differentiable, uniform noise is introduced to enable differentiable operations, as described in \cite{DBLP:conf/pcs/BalleLS16}. Thus, the latent space quantization is defined as:
\begin{align}
\hat{y}_{k|j}^{b} &=
\begin{cases}
     y_{k|j}^{b} + u, & u \sim \mathcal{U}[-0.5,0.5] \text{ during training} \\
     Q(y_{k|j}^{b}) & \text{ otherwise}
\end{cases} \nonumber \\
\text{where } &\mathcal{U} \text{ is the uniform noise and } j \in \{1,...,N\}
\end{align}
\vspace*{-1.0\baselineskip}
\subsection{Autoregressive model}
\vspace*{-0.5\baselineskip}

The auto-regressive probability model $f_{\psi}$, implemented as MLP aims to closely estimate the quantized latent distribution $p_{\psi}$. Since the distribution of each pixel in the latent space is conditioned by their neighbors, according to \cite{DBLP:conf/nips/MinnenBT18} the probability of the pixels is determined by a factorized model:
\begin{equation}
p_{\psi}(\mathbf{\hat{y}}_j)=\prod_{i,k}p_{\psi}(\hat{y}_{ik|j}|c_{ik|j})
\label{eq:factorizedmodel}
\end{equation}
where $\hat{y}_{ik|j}$ is the latent pixel at the position $i$ of level $k$ of description $j$ and $c_{ik|j} \in \mathbb{Z}^\mathcal{C}$ are the set of decoded neighboring pixels $\mathcal{C}$ of $\hat{y}_{ik|j}$ representing decoding context. The auto-regressive model $p_{\psi}$ uses the Laplace distribution as described in \cite{arxiv:ladune2023coolchic} to approximate the real probability of latent space and by using the factorized model equation (\ref{eq:factorizedmodel}), the rate for each description $\mathbf{\hat{y}_j}$ can be expressed as:
\begin{align}
    \nonumber R(\mathbf{\hat{y}}_j)& = -log_2(p_{\psi_j}(\mathbf{\hat{y}_j}))= -log_2\prod_{i,k}p_{\psi_j}(\hat{y}_{ik|j}|c_{ik|j})\\\vspace*{-0.75\baselineskip}
    &=-\sum_{i,k}log_2p_{\psi_j}(\hat{y}_{ik|j}|c_{ik|j})
    \label{eq:SumOfEntropy}
\end{align}
\vspace*{-1.5\baselineskip}
\subsection{Multiple description optimization}
\vspace*{-0.5\baselineskip}
The optimization process is divided into two distinct phases: training and post-training. 
The objective of the training phase is to update the model parameters \( \theta \) and \( \psi \), and to adapt the various latent spaces \(\{ \mathbf{y}_1, \ldots, \mathbf{y}_N \}\) to the dynamics of the target image. Its cost function is defined as:
\begin{equation}
    J_{t} = D_0(\mathbf{\hat{y}}_0) + \alpha\sum_{j=1}^{N}{D_j}(\mathbf{\hat{y}}_j) + \sum_{j=1}^{N} \lambda_{j}R(\mathbf{\hat{y}}_j)
    \label{eq:costfunctiontraining}
\end{equation}
where $\alpha \in [0,1]$ is the redundancy factor, $R(\mathbf{\hat{y}_j})$ denotes the rate as defined in equation (\ref{eq:SumOfEntropy}), $D_j$ is the side distortion, and $D_0$ is the central reconstruction distortion. The differences in distortion, represented by \(D_1, \ldots, D_N\), between the side reconstructions \(\mathbf{\hat{x}}_1, \ldots, \mathbf{\hat{x}}_N\) and the central reconstruction distortion \(D_0\), are dependent on the redundancy factor \(\alpha\).  The configuration of cost function (\ref{eq:costfunctiontraining}) pushes the Synthesis model to partition the image information into $N$ distinct descriptions and converges towards maintaining the lowest $D_0$ possible while accommodating different rates.
After training the network, the model parameters $\psi,\theta$ are represented as 32-bit values, but such precision is not necessary for transmission. Thus, in the post-training phase
the model parameters $\theta$ and $\psi$ are quantized according to equation (\ref{eq:scalarquatizer}), transforming them into $\hat{\theta}$ and $\hat{\psi}$, respectively. The quantization steps for $\hat{\theta}$ and $\hat{\psi}$ are optimized as outlined in \cite{le2023inrmdsqc} by minimizing the post-training cost function:
\vspace*{-0.5\baselineskip}
\begin{align}
\nonumber J_{p} =& D_0(\mathbf{\hat{y}}_0,\mathbf{\hat{\theta}},\mathbf{\hat{\psi}}) + \alpha\sum_{j=1}^{N}{D_j}(\mathbf{\hat{y}}_j,\hat{\theta},\hat{\psi}) \\ \vspace*{-0.75\baselineskip}
&+\sum_{j=1}^{N} \lambda_{j}(R(\mathbf{\hat{y}}_j, \hat{\theta}, \hat{\psi})+R(\hat{\theta})+R(\hat{\psi}))
\label{eq:posttrainingcostfunction}
\end{align}
Where, $R(\hat{\theta})$ and $R(\hat{\psi})$ represent the estimated rate utilizing a Laplace model.
\vspace*{-0.5\baselineskip}
\section{Entropy coder adapted to DNA}
\begin{figure}[!htp]
    \centering
    \includegraphics[width=0.9\linewidth]{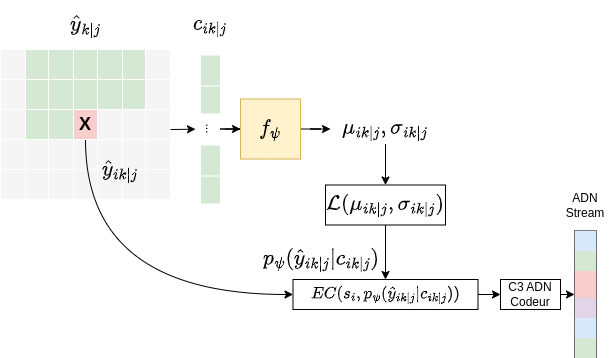}
    \caption{\footnotesize \textbf{Context Entropy Coding with C3 DNA coder:} In this example, the model uses 12 pixels, $c_{ik|j}$, to yield ${\mu_{ik|j}}$ and ${\sigma_{ik|j}}$, modeling a Laplacian distribution. The symbol probability is calculated, and an entropy decoder estimates the latent pixel, $\hat{y}_{ik|j}$, from a bitstream. The bitstream is then converted to quaternary code by using the C3 DNA coder.}
    \label{fig:CtxCoding}
    \vspace*{-1.0\baselineskip}
\end{figure}%

\vspace*{-0.5\baselineskip}
\subsection{Description coder: Range Transcoder}
\vspace*{-0.5\baselineskip}
In the binary case, the Range coder \cite{bamler2022constriction} has been used to entropy code the latent space in different MDC schemes.
Since the Range coder offers high performance at a very low entropy, we decided to adapt it to DNA by designing a transcoder that encodes its output into DNA.
The principle of context latent coding is depicted in Fig. \ref{fig:CtxCoding}. The encoded values from Range coder are then fed to the $C_3$ coder described in \cite{xavier_icip_2023}. 
In this paper, we introduced an arithmetic coder inspired by the JPEG 2000 MQ coder. This coder is based on a fixed-length code \( C_3 \) composed of 48 elements. Further inspired by the Run-length Limited (RLL) binary codes, it has been designed to prevent the occurrence of homopolymers, which are repetitions of the same nucleotide too many times consecutively. The ANS coder output will be represented in base 48. Its base 48 development will be encoded in DNA with the $C_3$ code.
\\ 
$C_3 = \lbrace AAT, AAC, AAG, ATA, ATC, ATG, ACA, \hspace{0.1cm}..., \\ \hspace{0.1cm} GCT, GCG, GGA, GGT, GGC\rbrace$
\\
$|C_3|=48$
\vspace*{-1\baselineskip}
\subsection{ARM and Synthesis Models coder: SFC4}
\vspace*{-0.5\baselineskip}

In \cite{SFC4}, we introduced a novel constrained quaternary entropy coder adapted to the biochemical constraints of DNA data storage, with increased performance over the state of the art Huffman/Goldman algorithm \cite{Goldman2013}. In \cite{le2023inrmdsqc}, the MLP can be modeled by a Laplace distribution, so the code-book is initialized with a frequency table following this Laplace model. After initialization, the SFC4 encoder will be used to encode all the parameters of the ARM and synthesis models, since they are necessary for decoding.
%
\begin{figure}[!htp]
    \centering    \includegraphics[width=0.98\linewidth]{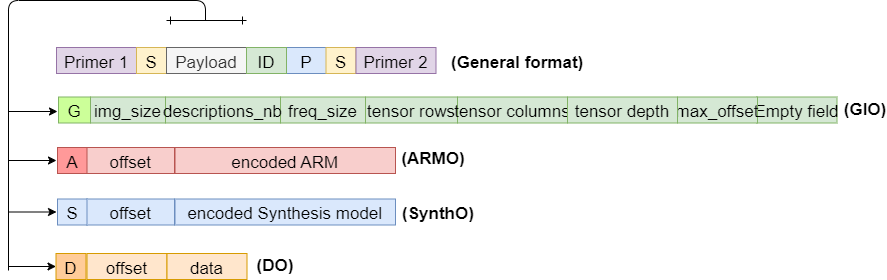}
    \caption{\footnotesize{\textbf{Design of the differents oligos format}. General format: The format remains consistent across all oligos, with the only variation occurring in the payload. "S" is the orientation nt, ID is the encoded file's label, P is a set of 4 parity nucleotides. GIO: General informations for the encoded image such as the image size, the number of descriptions and the coding dynamics. ARMO: Contains the weight and the bias of the ARM model. SynthO: Contains the weight and the bias of the Synthesis model. DO: Contains the latent spaces' pixels.}}
    \label{fig:formatting}
    \vspace*{-2.0\baselineskip}
\end{figure}
\section{Oligo structure \label{sec:bsstructure}}
\vspace*{-0.5\baselineskip}
\begin{figure*}[!htp]
\vspace*{-3\baselineskip}
    \centering
\begin{minipage}{0.45\textwidth}
    \centering
\includegraphics[width=0.81\linewidth]{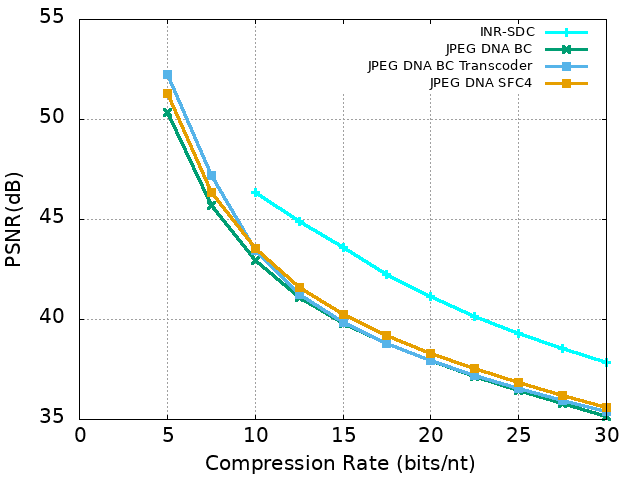}
\vspace*{-1.0\baselineskip}
    \captionof*{figure}{\footnotesize (A) SDC's Rate Distortion}

\end{minipage}
    \begin{minipage}{0.45\textwidth}
    \centering
\includegraphics[width=0.81\linewidth]{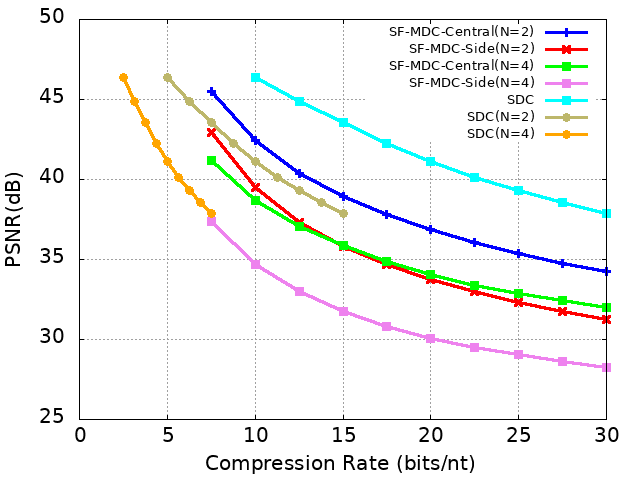}
 \vspace*{-1.0\baselineskip}
    \captionof*{figure}{\footnotesize (B) MDC's Rate Distortion}
\end{minipage}
\vspace*{-1\baselineskip}
\caption{\footnotesize (A) Average results over the kodak dataset. Our novel SDC coding scheme overperforms all the state of the art coders by at least 0.5 to 3 dB (JPEG DNA BC: \cite{Dimopoulou2021AJI}, JPEG DNA BC Transcoder: \cite{EPFL}, JPEG DNA SFC4: \cite{SFC4}), (B) Average result curve over kodak dataset, the MDC side curve is the mean curve across different descriptions. The benchmark is done with the following configuration: N number of descriptions with $N=\{2,4\}$, $\alpha=0.1$. The SDC ($N=\{2,4\}$) is its rate $\times N$, it is equivalent to the compression rate used with MDC ($N=\{2,4\}$), it allows us to compare SDC and MDC in terms of quality for the same rates.\label{fig:perf}}
    \centering
    \begin{minipage}{0.25\textwidth}
        \centering
        \includegraphics[width=0.90\linewidth]{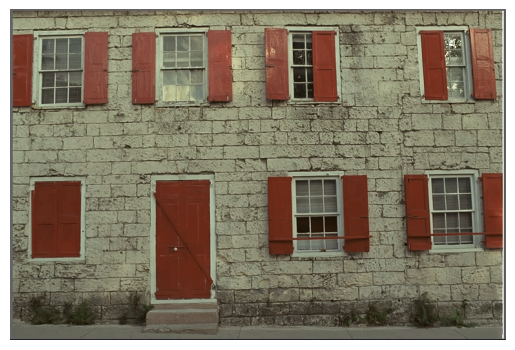}
        \vspace*{-1.0\baselineskip}
         \captionof*{figure}{\footnotesize (A1) Side description 1: 38.757 dB}
         \vspace*{-0.2\baselineskip}
    \end{minipage}%
    \begin{minipage}{0.25\textwidth}
        \centering
        \includegraphics[width=0.90\linewidth]{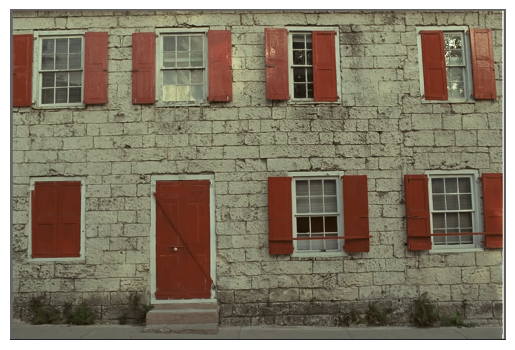}
        \vspace*{-1.0\baselineskip}
         \captionof*{figure}{\footnotesize (A2) Side description  2: 38.789 dB}
         \vspace*{-0.2\baselineskip}
    \end{minipage}
    \begin{minipage}{0.25\textwidth}
        \centering
        \includegraphics[width=0.90\linewidth]{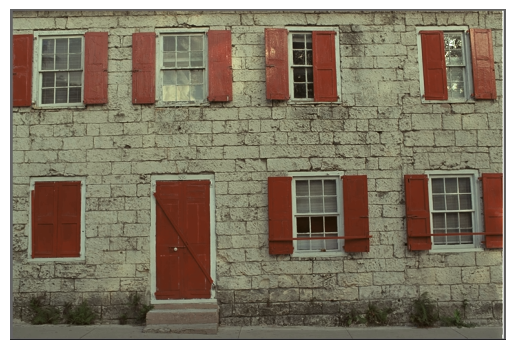}
        \vspace*{-1.0\baselineskip}
         \captionof*{figure}{\footnotesize (A3) Central description: 41.829 dB}
         \vspace*{-0.2\baselineskip}
    \end{minipage}
    \begin{minipage}{0.25\textwidth}
        \centering
        \includegraphics[width=0.90\linewidth]{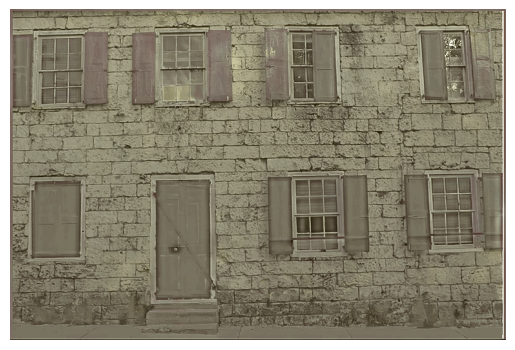}
        \vspace*{-1.0\baselineskip}
         \captionof*{figure}{\footnotesize (B1) Side description 1: 20.329 dB}
    \end{minipage}%
    \begin{minipage}{0.25\textwidth}
        \centering
        \includegraphics[width=0.90\linewidth]{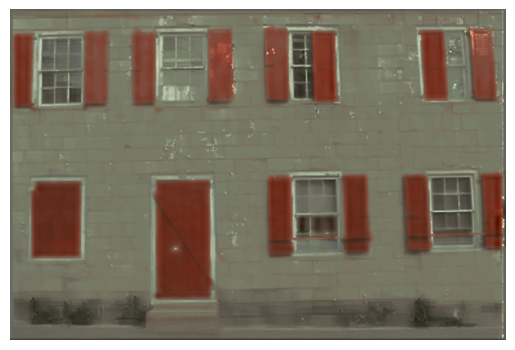}
        \vspace*{-1.0\baselineskip}
         \captionof*{figure}{\footnotesize (B2) Side description  2: 20.039 dB}
    \end{minipage}%
    \begin{minipage}{0.25\textwidth}
        \centering
        \includegraphics[width=0.90\linewidth]{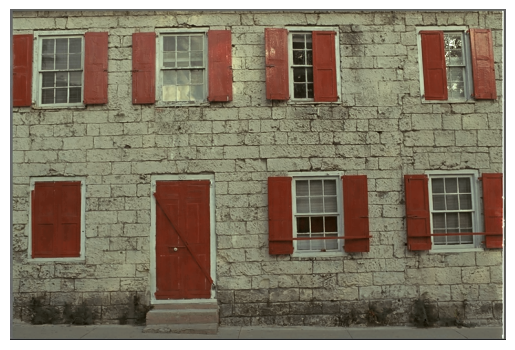}
        \vspace*{-1.0\baselineskip}
         \captionof*{figure}{\footnotesize (B3) Central description: 35.829 dB}
    \end{minipage}
\vspace*{-1.0\baselineskip}
    \caption{\footnotesize \textbf{Loss simulation on kodim01}: 
The image is encoded with two side description shown in (A1) and (A2). The central reconstruction computed from these side descriptions is shown in (A3).
Noise was then introduced (oligo loss), removing entire latent spaces from those side descriptions. (B1) and (B2) are respective visual results of this noise added to the side descriptions (A1) and (A2), and (B3) is the visual result on the central reconstruction computed from (B1) and (B2).}
\vspace*{-1.5\baselineskip}
\label{fig:visualisation}
\end{figure*}

DNA data storage requires the use of short strands called oligos, of length generally lying between 100 and 300 nts. In this work, we use oligos of length 200 nts. 
The decodability is ensured only if we manage to decode at least one of the descriptions, the Auto Regressive Model, and the synthesis model. In our design, we separated the different parts of the encoded data into separate oligos. Some oligos will encode the ARM model, some the synthesis, and other oligos will encode separate latent spaces, as presented in Fig. \ref{fig:formatting}. 
\vspace*{-1.0\baselineskip}
\section{Experiments}
\vspace*{-1.0\baselineskip}
In the following subsections, we are going to introduce comparative results from different DNA coding methods. All the coders presented here use oligos of length 200 to avoid generating side effects on one of the method's performance. The images used to conduct the test are extracted from the previously mentioned kodak dataset. The number of hierarchy levels used is six ($L=6$).
\vspace*{-1.0\baselineskip}
\subsection{Performance study}
\vspace*{-0.5\baselineskip}
The new SDC shows better performance over the state of the art image coding methods adapted to DNA as shown in Fig. \ref{fig:perf}(A). With this new method, we were able to show gains between 0.5 and 3 dB in terms of quality of reconstruction in comparison to the best previous method (JPEG DNA SFC4 Transcoder). The results have been computed and averaged on the kodak dataset.

To ensure the validity of SF-MDC, its performance at central reconstruction should neither surpass the upper limit of the SDC nor fall below the SDC at an \(N \times\) Rate. As shown in Fig. \ref{fig:perf} (B), with \(\alpha=0.1\), the solution approaches the upper bound limit of the single SDC as the rate increases, and never goes under the lower bound limit for different $N$. Besides, we observed that the compression rate increases with the number of descriptions used. On the other hand, increasing the number of description makes the coder more robust to noise.

\vspace*{-1\baselineskip}
\subsection{Noise robustness}
\vspace*{-0.5\baselineskip}

In this section, we simulate the loss for the MDC case $N=2$. As each latent space is entropy coded and independently decodable. Therefore, to analyze a typical case scenario, we drop three out of six latent spaces from each description, alternating between different levels of descriptions (Description 1: 77\% oligo loss, Description 2: 23\% oligo loss, and Central Description: 50\% oligo loss). The results have been computed on the image kodim01 of the kodak dataset previously mentioned. As observed in Fig. \ref{fig:visualisation}, the MDC demonstrates a high resilience capacity, maintaining a loss of only 5dB when losing a big part of the information contained in the different latent spaces.

\vspace*{-1\baselineskip}
\section{Conclusion}\vspace*{-1\baselineskip}

This work introduces an innovative DNA-based image codec that achieves substantial improvements in reconstruction quality when compared to existing DNA-based image codecs. On average, these improvements amount to 3 dB, with peak gains of up to 5 dB. These notable enhancements result from the utilization of the ARM, synthesis networks, and the DNA-adapted ANS coder, which deliver exceptional performance even at low entropy levels.

Furthermore, we present a Multiple Description Coder (MDC) capable of generating a variable number of descriptions. This MDC enhances the resilience of oligos to the noise inherent in DNA data storage channels. We also conducted experiments that involved introducing noise into the storage channel. The result shows that we only lost 5dB in the worst scenario.

In future works, we aim at building a noise model for the DNA data storage channel that could further improve the noise robustness of the  MDC.
\vspace*{-0.5\baselineskip}
{\footnotesize
\subsection*{Acknowledgement}
\vspace*{-0.5\baselineskip}
We would like to extend our gratitude to Dr. Eva Gil San Antonio for her innovative ideas, suggestions that improved the quality of the paper.
}\newpage
\bibliographystyle{IEEEbib}
{
\bibliography{refs}
}
\end{document}